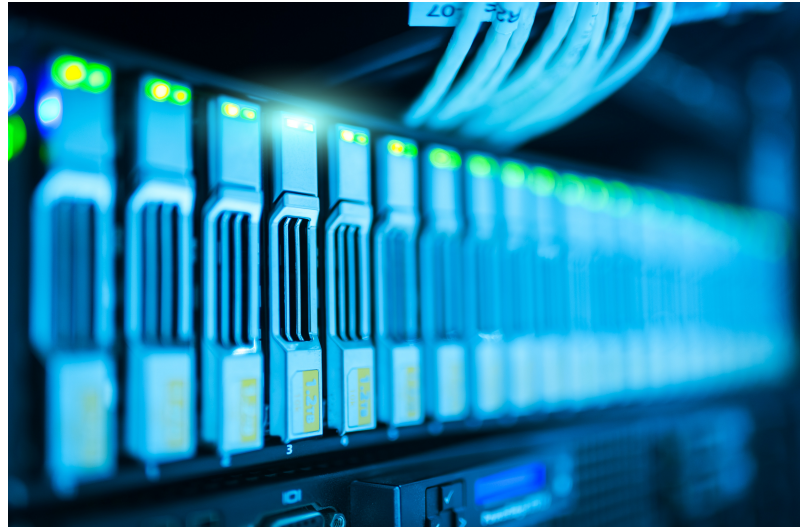

# Challenges in Network Management of Encrypted Traffic

**White paper by the EU-H2020 MAMI project (grant agreement No 688421)**

Mirja Kühlewind (ETH Zurich), Brian Trammell (ETH Zurich), Tobias Bühler (ETH Zurich), Gorry Fairhurst (University of Aberdeen), Vijay Gurbani (Illinois Institute of Technology)

The MAMI project held a Management and Measurement Summit (M3S)  in London on March 16, 2018. This meeting was an industry focused, invitation-only workshop with the goal to discuss challenges in network management that arise with the increased deployment of traffic encryption on all layers. Further information on the workshop can be found on the [MAMI project website](#).

This paper summarizes the challenges identified at the MAMI Management and Measurement Summit (M3S) for network management with the increased deployment of encrypted traffic based on a set of use cases and deployed techniques (for network monitoring, performance enhancing proxies, firewalling as well as network-supported DDoS protection and migration), and provides recommendations for future use cases and the development of new protocols and mechanisms. In summary, network architecture and protocol design efforts should

1. provide for independent measurability when observations may be contested,

2. support different security associations at different layers, and

3. replace transparent middleboxes with middlebox transparency in order to increase visibility, rebalance control and enable cooperation.

In recent years, the increase in public concerns about surveillance of and interference with Internet traffic have led to a rapidly expanding deployment of encryption to protect end-user privacy, mainly based on the use of HTTPS [RFC2818] and thereby TLS [RFC8446]. This evolution provides a step-function increase to the protection of a user's privacy and data. Still all information that is radiated from an endpoint may contain user-specific information and should be evaluated for its privacy implications.

Further, encryption can provide the sender of a packet with clear control of which data and information can be accessed by which entity, e.g., an on-path entity, at a specific protocol layer. Therefore, encryption also helps to address the ossification problem that has made it more and more difficult to deploy new protocols and protocol extensions using the Internet. The extension of the encryption boundary into the transport layer, beyond what TLS over TCP protects, as we see with QUIC, is the next step to protect even more information at even lower layers.

However, this move to enforce protection of lower-layer headers with encryption will also necessitate a change to the model that is often used today to provide in-network functionality. This functionality includes simple firewalling, DDoS protection, or performing enhancing services that may be performed with or without the explicit approval of one or both communication endpoints. More generally, these functions can be split into five classes regarding data access and exchange requirements:

- Explicit exposure of information for network consumption (where communication is end-to-end integrity protected, and no trust relationship with midpoint required);

- Split-transport proxying (where a higher layer is encrypted end-to-end, and lower layers are optionally encrypted with the midpoint as a communication end);

- Network to endpoint signaling (where the length of a "scratch" space can be end-to-end integrity protected, depending on the encryption this optionally requires a trust relationship);

- Full visibility, for content inspection without modification (where a trust relationship is required with at least one endpoint);

- Content modification (where a trust relationship between both endpoints and the midpoint is required).

> "What do you do when the endpoint can't trust the network? And the network can't trust the endpoint?"

Based on this classification we can say that the motivation for encrypting data at different layers can be very different. If two layers are encrypted for the same reason, their security associations can share the same encryption envelope. However, if they encrypt for different reasons (e.g., one for application privacy, another for transport operational correctness), then there is an opportunity to encrypt them differently, or to even use different security associations for each layer with potentially even different communication endpoints. Otherwise, the functions listed above can become either unnecessarily restrictive or more information than necessary may be exposed. Instead, we should preserve the diversity of solutions and the ability of users to choose at each layer.

Even if different encryption envelopes are used to separate information accordingly, it should be noted, that it does not make sense for midpoints within the network, which do content inspection for security support, to provide partial visibility, e.g., visibility of data at only some of the layers. Encryption of the data at other layers may then hide the information to be monitored, therefore these functions will always need to see all data.

An IAB document [ID.wire-image] further describes implications to the integrity and confidentiality protection of the *wire image* of a network protocol, i.e. the information content derivable from passive observation of packets belonging to a communication.

**Impact on protocol evolution**

Internet transport has always interacted with network operations to find and configure suitable network paths, and to ensure the network as a whole works well for customers and operators. Building any transport system requires careful engineering to achieve a good (and affordable) result. This holds for the Internet just as it does for physical transport systems such as the London Tube. The Tube system doesn't "just work"; it requires constant maintenance and engineering support to deliver reliable and appropriate service. Transport protocol design needs to consider this deployment aspect as well.

> "Users may follow the law, attackers probably will not. We need models when analyzing security with a higher granularity than "an attacker" and "not an attacker" -- there are different users of networked systems, each of which have different incentives."

Transport layer header information has for a long time been used to support network operations, and the network operations community has in the past benefited from understanding the pattern and requirements of traffic passing through their networks, both at an aggregate and flow level [ID.encrypt]. For example, visibility of transport header information has allowed toolchains to be used and maintained that help understand customer quality of experience, detect performance or configuration issues, identify the root cause of issues and interactions, monitor the impact of changes in infrastructure/configuration and in response to changes in application traffic patterns.

> "Some level of encryption will likely form the basis of future protocol designs."

New protocols should be designed by making appropriate use of authentication and encryption. Authentication protects the integrity of transport information and prevents undetected changes to the header information. Encryption provides confidentiality of the transport protocol header, preventing any on-path function from using information therein. However, increasing use of encryption will also reduce the ability of operators and researchers to measure compliance with transport specifications and interactions between protocol features and configurations to better understand the operation of the Internet as a whole.

In the past, measurement data and experiments conducted using in-situ equipment have helped to understand design trade-offs, informed selection of appropriate mechanisms, and helped to ensure a safe, reliable, and robust Internet. The sort of measurements that are impacted by increasing confidentiality are not related to application development, where a test node can enable specific functions, e.g., share a key or make data public, but will impact any measurement at scale, and experiments that look at complex interactions between application flows, configuration, policy and coexisting traffic.

This leads to questions about whether this trend will have an impact on transport evolution, which has in the past benefited from measurements within the network and insight provided (often indirectly) by the network operations community: Could the long-term implications of this trend even change the motivation for designing common protocol specifications and the way in which research data is able to support standardisation? There is also a question of who bears the resulting costs that encryption could place upon the research community. Going forward, it will be important to understand that there are real risks to future innovation from these impediments when encryption is applied without considering the impacts on operations, research & development, and regulation.

# Network Management Use Cases

We considered four major use cases for network management: network measurement for monitoring and troubleshooting, protection mechanisms such as firewalls, performance-enhancing proxies, and in-network DDoS mitigation. More detailed information on each can be found online in the respective workshop slides.

## Network Performance Monitoring

Performance measurements can support a variety of activities: response to customer issues, monitoring when deploying new features, detection or response to anomalies, understanding where faults are located, or to support planning for investment and sustainability.

A wide range of performance evaluation tools and measurement techniques has evolved over time for basic network monitoring. Many operators have added their own solutions over time, e.g., to measure RTT and throughput, that often rely on at least transport header information that is visible in clear with TCP. These tools are now well-understood and most tools are now trusted by operators, because they need to be relied upon them.

While it is good to see an increase in the pace of evolution, this comes either at additional cost in development of toolchains and operational procedures, or alternatively with a loss of detail that will change how the network service is offered. Some existing tools will not work with encrypted traffic. Other techniques, such as load-balancing, often rely on just the 5 tuple to be visible. These should still work with transport encryption, e.g. QUIC [ID.QUIC], but might also have limited usefulness for more complex scenarios.

> "Sometimes the operator provides a bit pipe, then you care about bit pipe measurements. Sometimes you are a service provider, then the service-level metrics are important. This is not just perception of the role... lots of people think the network should be a dumb pipe, and that running the dumb pipe is easy. This is fundamentally a struggle over the value... This is not an obvious trade off."

Services that operate on a per-user basis (e.g., mobile user equipment), can currently utilise measurements to augment information about subscribers, to add context about the service being used (e.g., to support Quality of Experience, QoE, evaluation), or to detect issues with specific parts of the infrastructure. For example, in mobile networks, flow-level metrics can be collected in the network within a GPRS Tunneling Protocol (GTP) tunnel, which allows measurements to follow the customer, and provides the identification necessary to associate measurements with specific equipment. Passive measurement can also be used to take a small number of samples. For example, one use could measure every 5 minutes, aggregating the statistics at the cell side, to observe throughput, half-RTT to track latency changes and infer congestion through queueing delay, or measure retransmissions to infer loss. These metrics can be correlated with layer 2 impacts to provide evidence of subtle configuration issues with equipment. A key point is that measurement data of different granularity and captured at various vantage points throughout the network is necessary to support effective operation of the network.

> "Operators would like to know the speed experienced, whether there was congestion, where there is any "bad connectivity", the achieved flow rate and jitter observed."

Measurements can help understand the impact of traffic. During peak evening hours, overall network congestion typically rises, with a correspondingly higher latency and higher retransmission rates. This is currently most noticeable on 3G networks, as these are often more congested and is less so on 4G networks. Drilling down by probe and network equipment provides an easy identification of the most congested network paths, and indication of the usage patterns that result in this. This method of analysis has repeatedly allowed network management teams to isolate network issues that typically do not show in the KPIs of individual routers or network hardware.

Passive measurements provide important performance data, but it is difficult to judge network speeds available to customers from passive measurements alone. A lot of user activity simply does not fully utilize the available speed. However, active measurement would be counterproductive for the busiest paths within networks, because these are precisely the paths where one does not want to add more traffic. Instead, it is common to look at customer satisfaction with the available bandwidth using passive monitoring. For example, a probe can sample the observed TCP goodput by region. A probability distribution function of the peak speeds can be used to predict where customers may be less satisfied.

If a customer does happen to complain, measurement data helps with troubleshooting, e.g., to understand whether their complaint is likely to relate to network congestion issues. The peak speed probability distribution over time can also be an indication of the average speed experienced, but this is not necessarily so. Correlating inflection points with customer survey results helps to identify areas where network capacity should be improved (i.e., where subjective ratings tip from Good to Poor for regions where customers experience less than 70%-80% of peak speed).

Encryption can hide measurement of the speed or performance actually achieved by a user. Many current passive measurement methods depend on seeing sequence numbers and ACKs, to follow the window size and make a judgement what determines the current window limitation: the network based on bandwidth limits or congestion; or the application above the transport that does have more data to send; or the transport's flow control itself. It is not a general solution to simply reduce the available toolset for network monitoring to only use active measurements, because active measurements do not observe the actual performance as experienced by the user. This will become more difficult when traffic uses encrypted transports, such as QUIC. Future tools could use methods such as the spin bit, which is a simple explicit signal for passive estimation of latency. Such an approach of simple explicit signalling could also be developed to measure loss and congestion, and thereby estimate the evolution of the window size over time.

*Example*: **Over-The-Top (OTT) Video Analysis**

One important example that currently relies on this information is measurement of the expected video quality across a particular network segment. Current tools exist to interpret the media rate of a streaming video session and to understand the quality of experience using a model based on subjective data by passively observing traffic. Such a model relies on key input information from the traffic such as content type, hostname, media container metadata, payload size, and transport dynamics to infer the network impact. An encrypted transport would hide application as well as transport information, such as the application payload size, sequence and acknowledgement numbers, retransmission. Also currently, DNS information can be used to augment these kind of measured data, however, the use of encrypted DNS would stop the use of these kind of information. Lack of quality information will not only provide challenges for the support of video traffic in general but can also be a barrier to entry for small content providers as the network may be optimized for large content providers only (in directly cooperation with them).

> "Applying passive measurement to generate quality of experience metrics in the absence of transport information would require heuristics to infer the dynamics based on traffic patterns, with resulting assumptions about what the application may be trying to do, and what the implications are for the applications/network. These assumptions are necessarily brittle, and lead to lower-quality insights about user-experienced performance."

Of course, some performance monitoring can also be done with independent QoS measurements (verification) once everything is encrypted (active measurements). This sort of measurement tool is often even appreciated by enterprises, as it can provide more detailed insights in the functioning and problems of a specific application. However, such an approach has higher costs and requires additional bandwidth resources and is therefore often only used for some dedicated services in addition to network monitoring.

Further, encryption can hide application decisions. For example, encrypting control traffic prevents measurements providing insight into the multiplexing part of a QUIC session, as one QUIC connection can carry multiple independent application services (e.g., audio and video, or additional control data). Encryption intentionally makes it more difficult to infer internals of the application design, but also hides likely interaction of different application services with changes in network conditions. This can make it impossible to know why the application traffic pattern changes in response to changes in the network. Such changes can occur very suddenly, e.g., when a new feature is enabled in an application (e.g. when a new version of QUIC is activated in Chrome browsers), and can have unexpected interactions with existing deployed equipment configurations in the network, requiring changes to the network configuration.

**Reducing visibility can impact industry competition**

Looking forward, it is also important to realise that there are various communities seeking to understand network performance and issues. This includes not only operators, but researchers and developers wishing to use the network, information about whether customers receive the advertised service that they have subscribed to, and regulators seeking to see the bigger picture. Increased use of encryption will also have an impact on service verification.

> "An inability to independently measure performance metrics leads to a business disadvantage -- if the media companies only cooperate on performance metrics with larger operators, smaller operators will be placed at a disadvantage. A danger emerges that favours companies that have large infrastructures."

The goal of a given performance measurement activity determines the set of metrics of interest. Even within an operator, there is a diversity of such goals. The faults reported also vary. For example, the first answer from a network operator asking about a problem may be, "the problem is with your network". This means any operator really has to prove there is no problem in their own network segment, not just that packets are sent along the service.

A balanced understanding of how the network is performing (or not) is not possible if only one party has access to these metrics. It becomes impossible to verify the applicability and veracity of those metrics. ***Open measurability at different places along the network path is therefore of the utmost importance*** in cases where there may be disagreement among parties about the parameters of the service provided. There are additional pressures on this: Some people have access to much more information than other people. This analytic data is important enough to people that it can be sold to those willing to pay.

## Performance enhancing proxies

An assumption of the Internet architecture is that any two cooperating endpoints can deploy a new application or transport layer protocol. However, the present Internet adds a set of third parties, middleboxes on the path between the endpoints. Many of these boxes implement in-network functions to improve performance of the transport when crossing low-performance links (high-loss, high-latency) or highly asymmetric links (narrow upstream bandwidth) [RFC3449]. Examples include, but are not limited to, satellite, mobile, and wireless access networks. To address this challenge Performance Enhancing Proxies (PEPs) [RFC3135] are put on one or both ends of the low-quality/asymmetric link. Today, these proxies may either intercept the transport connection completely ("back to back termination"), or perform various of "tricks" with TCP headers such as ACK manipulation (suppression, reconstruction, compaction), receive window size manipulation, and sometimes header compression.

While these middleboxes can address the problem of link asymmetry or packet loss, which is hard to address solely on an end-to-end bases, these approaches are also knowingly problematic as they make assumptions about the traffic flowing through them, and thereby impose limitation on protocol evolution: "*All these middleboxes optimize current applications at the expense of future applications. In effect, future applications will often need to behave in a similar fashion to existing ones, in order to increase the chances of successful deployment. Further, the precise behavior of all these middleboxes is not clearly specified, and implementation errors make matters worse, raising the bar for the deployment of new technologies*" [RFC6182].

This problem can be addressed by header encryption, which makes it impossible for on-path middleboxes to modify transport headers. This has a tradeoff in that any middlebox PEP functions are completely inhibited. For example, it is impossible to deploy an ACK-suppression scheme for QUIC, because acknowledgments in QUIC are encrypted. However, making PEPs go away does not make the problems they address go away.

In moving to encrypted transports, it is either necessary for the endpoints to handle diverse range of performance issues that arise from challenging network paths (e.g., highly asymmetric performance at network boundaries, intermittent reordering, loss, widely varying capacity, etc) or it is necessary to design "hooks" into these protocols in order to allow for limited PEP functionality. For example, QUIC could eventually be extended to support transport-split proxies while still maintaining end-to-end encryption for payload and other control information.

Alternately, an approach using partial tunnels (e.g. semantically equivalent to HTTP CONNECT, but used for building UDP tunnels instead of TCP) and leveraging higher-layer awareness of performance diversity at network boundaries could potentially address some of these issues.

*Example*: HTTP proxy for network multicast support

High resolution (live) video content is seen by a large number of viewers, therefore there are opportunities for delivery over multicast in the network. This can be very beneficial for saving bandwidth. In such a setup, multicast is sometimes already used from content provider to a home gateway where the home gateway is an HTTP proxy and multicast-to-HTTP/S-unicast transport translator. In such an example setup, the content playback function fetches the manifest file from content hosting and uses its content to retrieve the media. The gateway also acts as reverse proxy (with DNS interception) and therefore requires a certificate (but does not have a static IP address, so certificates should have a DNS name). Moreover, the gateway would need to change the manifest (URLs).

In the future, such a gateway could also in future be co-located with the home setup box/router but this would introduce a major security challenge as this requires to place the content provider certificates and manifests on the gateway.

Removing the separation between the security and the application layer is problematic

More generally, there are two types of interception: passive without modification and therefore termination is optional, and active that need to terminate the connection to modify the traffic. Today, traffic interception needs to consider three layers: the transport (TCP/IP), the security (TLS), and the application (HTTP) layers. The interception type may be different for each of the different layers. For example, a transport layer active intercept may need to terminate the transport connection to ensure that the correct options are chosen, but can be protocol agnostic to any higher layer. However, active application layer interception would usually need to also intercept for the lower layers/terminate their connection as well. Even approaches attempting to avoid the necessity of full termination might have to partially fake the lower-layer protocol dynamics to avoid failing to split the connection, because a full handshake response from the server is a necessary input to a policy decision.

The trend to remove the separation between the security layer and the application layer, while improving performance and security, is problematic for use of transport layer intercept devices. Currently these devices are able to inspect the transport layer while leaving the application layer encrypted, either by choice (i.e., following the principle of least access) or for legal reasons (i.e., jurisdictions that make a distinction between communications metadata and payload.)

## Protection and Firewalls

A middlebox that is set up to block malicious traffic needs to be able to classify the traffic as malicious or benign in order to make a policy decision. E.g. for TLS-encrypted traffic, the Service Name Indicator (SNI) is often used as an early classifier: benign connections can be quickly ignored if it has been validated that it is possible rely upon the SNI (i.e., the SNI in a ClientHello TLS message must match a Subject Alternate Name (SAN) in the server certificate). Unfortunately, TLS 1.3 prevents transport layer passive intercept devices from validating the SNI, which would requiring these devices to intercept all sessions. Further, if the SNI is encrypted, as currently proposed, [ID.tls-sni-encryption], none of this information would be available at all. Therefore an unintended consequence of SNI encryption is that all traffic needs to be inspected, not just traffic for which an early policy decision can be made. This has cost and complexity implications.

Further, as attacks on middleboxes become more likely, a requirement to intercept all traffic provides an additional attack vector. It is therefore important to understand and evaluate the benefits as well as costs of these kind of encryption decisions. As these security functions are often required by regulation and/or law, SNI encryption will not eliminate the deployment of these interception functions. It will simply make their use more costly. Alternatively, if proxies would no longer be able to inspect any traffic at all, these functions would need to be moved to the endpoints, at commensurately even higher cost of operation.

*Example*: Policy of blocking uncharacterised traffic in enterprise networks

Enterprises and smaller companies often have different needs and perspectives to network operators. They often just block protocols and functions that do not match the security policy that they implement (e.g. UDP blocking). However, they are also often unaware of industry and user trends. New traffic patterns, that suddenly emerge and may bypass existing security boxes/firewalls, would be in violation of the desired security policy of the organization. Therefore a common immediate reaction is to simply block all unknown traffic as it is observed (e.g., by port number, or by stripping unknown extensions). This general approach seeks to always "downgrade" to a standard/known behavior, thereby preventing the benefit of new features, until the in-network functions, such as firewalls, support them.

> "If you can't understand a flow, then you should block the traffic. New techniques, even if they are possible, always require tooling support. This leads to a temporary incentive to block unknown traffic, until the tooling supports it."

As such, a policy of blocking uncharacterised traffic poses future problems by ossifying the network. New protocols present a need to change: Multipath TCP is a problem as it leads to an incomplete view of a flow at security devices that assume unipath routing. A simple fix to make these security devices multipath-aware, faces the problem that it is fundamentally difficult to monitor asymmetric and multipath routing. This is similar to early problems when some security devices had to deal with packet fragmentation, the common fix, defragment all packets, is fundamentally less scalable than discarding fragments and fixing this issue elsewhere.

Further, when more content providers begin to deploy and use QUIC, the downgrade reaction of simply blocking QUIC will work without much perceived loss of functionality. This is because current efforts implement a fallback to TCP. For example, the Chrome browser currently makes decisions of whether to use QUIC vs. TCP based on various information, including computer hardware capabilities. This masks a QUIC failure, and downgrading is therefore a low-risk operation. However, if developers later ship browsers that do not fall back to TCP, or new applications are deployed that only work over QUIC, then these applications will become inoperable.

**We need to replace invisible interception with visible interception**

There is also a marketing consideration here: It would be nice if devices could be "fire and forget" but the reality is that this sort of device needs to be actively maintained to accommodate new protocol features. For example, on the one hand a security policy may wish to, or need to, ban Chrome or QUIC. On the other hand, employees want to use this due to advantages they perceive from doing so (e.g., speed, convenience, security).

The half-life of a broken or out-of-date function depends on the layer. Upper layer features, security features, etc, change quickly. However, complexity is also a consideration: Multi-layer full-stack proxies have a much longer testing and redeployment cycle. This is often not compatible with a desire for a fast upgrade cycle. The trend in removing layer separation in general, but specifically to security functions, makes the support of middlebox even more difficult and slower.

In the future, some environments could evolve protection services that migrate to the cloud to prevent direct access to remote endpoints except via the cloud service. Some client endpoint information could then even come directly from the cloud infrastructure. The local firewall would then only need to decide about the traffic to the trusted cloud. Also pulling logs directly from cloud services becomes more scalable since the flow information is always in one place. There are other reasons beyond visibility for moving security-related services to cloud deployments, so this could anyway happen as a side effect of other industry trends.

> "We should not assume what the endpoint wants to do, instead let the endpoint tell you. If you are too transparent, it is hard to tell why something is breaking. Is the blockage intentional, or unintentional… So let's try to make this less transparent!"

Moreover, middleboxes are not just a single-organization operation. Currently, middleboxes are often configured statically, but increasing composability of middleboxes requires inter-middlebox communication and coordination. In the future, a chain of trust among the various boxes within and across administrative domains will become necessary. There is a need to build a mechanism for middleboxes to discover each other, and to determine whether they trust the path. Similarly, a mechanism for endpoints to have enough information to decide whether they can trust the middleboxes to use the offered in-network function. That means we need to replace invisible interception with visible interception and give the client (and server) more insight through which middleboxes their traffic is going.

# DDoS Protection and Mitigation

Distributed Denial of Service (DDos) protection is necessary for an operational network. This includes methods to detect potential DoS traffic and to divert this to dedicated scrubbing centres that can be used to filter out DDoS traffic while passing as much legitimate traffic as possible, based on various features extracted from the input traffic as well as reputation and customer policy rules. Detecting potential DoS traffic needs to be done locally and requires some visibility of the traffic itself. Detecting unusual behavior of a given traffic aggregate is relatively easy while the classification of traffic as legitimate or suspect, which is needed to make the actual delegation decision, is more difficult.

It is even more challenging to mitigate distributed denial of service attacks in networks. Effective real-time detection of attack traffic in large-scale networks is generally achieved by combining multiple techniques in distinct phases. After the first phase, which detects attack traffic and segments it into candidate attack traffic and normal traffic, the former category of traffic is then presented to the second stage discriminators that apply more complex models to uncover attacks from a global perspective (i.e., whether the same attack is being launched at different ingress points into the network). The presence of pervasive encryption in modern leads makes all of these phases more challenging. In the absence of cleartext payload, at best the headers of a packet can be used for extracting flows and constructing complex models from the flows and the content of individual headers. Even this level of visibility of the headers is hindered when the entire packet is encrypted, as when using IPSec. In the worst case when the entire packet is encrypted, meta information like the packet length and inter-arrival times can be used as features to fit an appropriate model.

**Building learning models for encrypted traffic becomes more challenging**

More generally, encrypted traffic makes it challenging to operate real-time analytic environments, that can be used for DDoS detection and other troubleshooting and monitoring functions in a large-scale network. The traditional layering model of the analytic environments did not pay much attention to security; the simplifying assumption was that cleartext data was captured (perhaps through port mirroring) and then made available to platforms where relevant portions of the data were extracted to fit a model. Pervasive end-to-end encryption renders this model obsolete.

Large-scale analytic environments support aggregators (intermediaries) that can summarise data and forward only the summaries to the analytic platforms. These aggregators would not be able to extract the data to summarise it, if the data had been end-to-end encrypted.

Building learning models for encrypted traffic becomes therefore more challenging. One way to mitigate this is to design analytic systems such that the data collectors (perhaps executing on remote hosts) and the models (executing on the analytic platforms) can be distributed and itself communicate over a secure channel. Aggregators, playing the role of intermediaries, would need access to certain portions of the data and prevalent techniques to allow intermediaries limited access to the data could be used in such cases.

# Conclusions and Recommendations

While the increasing deployment of encryption up and down the network stack has undeniable benefits for privacy, security, and the evolvability of networks, it is important to consider and evaluate the impacts of this increased use of encryption. The considerations are not restricted to best practices in applications development, but span a range of operational and other issues that have come to rely upon observation of transport header information to deliver the services that they provide.

The impact of the increasing encryption of network traffic is on the agenda of almost every forum covering standardization and operational practice for the Internet and Internet-connected networks: including the IETF, ETSI, the ITU-T, and 3GPP. In the IETF in particular, the discussion has focused on strong end-to-end encryption as the only way to enable privacy and security for end users, drawing on a long organizational tradition [RFC1984] [RFC2804] of support for the end-to-end principle.

There is, however, an argument for the middle. In some environments, the network operator may not be able fully rely on every single user and device to implement appropriate security measures. This is a common condition today in enterprise networks. In such scenarios, locating the security services in the network can scale far better than a solution focused on the endpoint. The tradeoff here is scalability against the additional risk that comes with a prominent attack surface presented by centralised in-network security functions.

Based on discussions at M3S, we make the following recommendations for network and internetwork architecture, the design and deployment of in-network devices, and the design and deployment of future Internet transport and application protocols:

## 1. Provide for independent measurability when observations may be contested

One implication of the widespread deployment of cryptography down the stack, as exemplified by QUIC, is the reduction of observability and measurability of Internet traffic. Information that was formerly radiated off of transport protocols is now easily available only to the endpoints. This places endpoints at an advantage with respect to observers in the middle. While this is the whole point of increasing deployment of encryption, it presents particular problems in cases where there is some inherent conflict between on-path observers and endpoints: for example, when each blames the other for a given performance issue.

In these cases, *independent (third party) verification and observability of network performance is essential*. This does not mean a reversion to cleartext to support existing passive measurement techniques; however, future protocol development should provide hooks for verifiable measurability, to support this independent measurement function for metrics and measurements likely to be in dispute.

## 2. Support different security associations at different layers for different purposes

Integrating the transport and security layer, as QUIC does, has many benefits. Among these are combining transport and cryptographic handshakes to provide low-latency connection establishment and resumption; increased deployment of confidentiality and integrity protection because this is no longer an "optional" feature; and resistance to ossification that supports further protocol evolution. However, integrating protocol layers, as in QUIC, has the disadvantage of binding the security association used to protect the transport protocol with that used to protect application payload.

While selective exposure of information is inappropriate for many in-network observation functions, particularly those concerned with security operations, it can be useful in building back-to-back transport proxies. Designing future integrated transport protocols to **support multiple concurrent** **security associations, each protecting a different part of the stack, when negotiated by the endpoints** would restore some of the flexibility of the current arrangement of TCP over TLS, while still retaining the advantages above.

## 3. Replace transparent middleboxes with middlebox transparency

Many of the problems caused by the widespread deployment of in-network functions, regardless of their purpose, arise because current in-network function designs have attempted to be as transparent as possible, in the sense of "invisible to the endpoints". This makes it easier to incrementally deploy middleboxes, because no change is required at the endpoint, and no traffic is disrupted providing that endpoint protocol implementations do not violate the assumptions made by the in-network function.

Increasing encryption requires a different approach, by restoring control over the entities with which an endpoint communicates. This suggests an architectural way forward, by completing the transfer of power to the endpoints, even when middleboxes are present. How this is realised will depend on the use case, sometime this will be the user's own policy decision, in some cases (e.g., a school, an enterprise, a shared computer), the security policy will be controlled by another party - but still needs to be known by a user. *__Middlebox interaction with network traffic must be explicit, visible to the endpoints, and only performed when consent is granted by the security policy for both of the endpoints.__*

This work is partially supported by the European Commission under Horizon 2020 grant agreement no. 688421 Measurement and Architecture for a Middleboxed Internet (MAMI), and by the Swiss State Secretariat for Education, Research, and Innovation under contract no. 15.0268. This support does not imply endorsement.